# Possible persistence of multiferroic order down to bilayer limit of van der Waals material NiI$_2$

*Hwiin Ju*[1,†], *Youjin Lee*[2,3,4,†], *Kwang-Tak Kim*[5], *In Hyeok Choi*[1], *Chang Jae Roh*[1], *Suhan Son*[2,3,4], *Pyeongjae Park*[2,3,4], *Jae Ha Kim*[6], *Taek Sun Jung*[6], *Jae Hoon Kim*[6], *Kee Hoon Kim*[4,5], *Je-Geun Park*[2,3,4,\*], *and Jong Seok Lee*[1,\*]

[1]Department of Physics and Photon Science, Gwangju Institute of Science and Technology (GIST), Gwangju 61005, Republic of Korea

[2]Center for Quantum Materials, Seoul National University, Seoul 08826, Republic of Korea

[3]Department of Physics and Astronomy, Seoul National University (SNU), Seoul 08826, Republic of Korea

[4]Institute of Applied Physics, Seoul National University, Seoul 08826, Republic of Korea

[5]Center for Novel States of Complex Materials Research, Department of Physics and Astronomy, Seoul National University, Seoul 08826, Republic of Korea

[6]Department of Physics, Yonsei University, Seoul 03722, Republic of Korea

ABSTRACT   Realizing a state of matter in two dimensions has repeatedly proven a novel route of discovering new physical phenomena. Van der Waals (vdW) materials have been at the center of these now extensive research activities. They offer a natural way of producing a monolayer of matter simply by mechanical exfoliation. This work demonstrates that the possible multiferroic state with coexisting antiferromagnetic and ferroelectric orders possibly persists down to the bilayer flake of NiI$_2$. By exploiting the optical second-harmonic generation technique, both magnitude and direction of the ferroelectric order, arising from the cycloidal spin order, are successfully traced. The possible multiferroic state's transition temperature decreases from 58 K for the bulk to about 20 K for the bilayer. Our observation will spur extensive efforts to demonstrate multi-functionality in vdW materials, which have been tried mostly by using heterostructures of singly ferroic ones until now.

Van der Waals (vdW) materials have attracted much attention as they host numerous novel properties even in the atomic thickness limit, with a recent focus on magnetic vdW materials.[1,2] For example, Ising-type magnetic orders are maintained down to the monolayer thickness for antiferromagnetic $FePS_3$,[3] and also for ferromagnetic $CrI_3$ and $Fe_3GeTe_2$.[4,5] Although magnetic orders are usually unstable for two-dimensional systems with weak or no magnetic anisotropy, long-range antiferromagnetic (AFM) orders of XY- and Heisenberg-type are observed down to few atomic layers of $NiPS_3$ and $MnPS_3$, respectively.[6-8] Also, the ferroelectric (FE) state in the monolayer limit was experimentally confirmed in $MoTe_2$, $In_2Se_3$, SnS, and several others, even at room temperature.[9-11] Along with a massive amount of technical interest in observing and controlling singly ferroic states, heterostructures are expected to provide a valuable platform for realizing multi-functional nanoscale devices. This naturally led many researchers to examine ferroic orders' proximity or coupling effects in heterostructures of vdW materials.[12-17] In particular, heterostructures of ferromagnetic and ferroelectric vdW materials were theoretically investigated. It was demonstrated that the magnetic state could be modulated by the change in the electric polarization and vice versa.[15-17]

It is natural to raise a question; is there a vdW material hosting both electric and magnetic orders simultaneously? In particular, does such a multiferroic (MF) state survive in the atomic thickness limit? In this work, we pay attention to $NiI_2$, of which multiferroicity is fully characterized in the bulk state.[18,19] We demonstrate that the ferroelectric-antiferromagnetic coupling is preserved down to the atomically thin limit of the bilayer flake. Ni ions' triangular lattice with no uniaxial crystalline anisotropy hosts a helical spin order at 58 K for the bulk $NiI_2$,[18,19] and the spin-order-driven spontaneous polarization arises simultaneously.[19] We trace the FE order parameter in $NiI_2$ utilizing the optical second-harmonic generation technique, which can probe the polarization magnitude and the polar axis. We found that the transition temperature shows a strong thickness-dependence, but it remains finite as about 20 K for the bilayer $NiI_2$. In particular, multi-domain states are manifested to have polarization axes lying along three crystallographic axes dictated by the Heisenberg spin Hamiltonian of the triangular system. We expect that this $NiI_2$ can enrich the library of vdW materials, which can be exploited to design and develop multi-functional nanoscale devices using vdW heterostructures.

$NiI_2$ has a $CdCl_2$-type crystalline structure, as depicted in Figure 1a. $NiI_6$ octahedra share the edge, and they form a two-dimensional triangular lattice with a van der Waals bonding between adjacent layers. $Ni^{2+}$ ions have the $S=1$ spin, and their magnetic interactions work effectively up to next-next nearest neighbors.[18] As temperature ($T$) decreases, magnetic transitions occur successively to the collinear AFM at 76 K and the helimagnetic state (Figure 1b) at 58 K, of which signs are indicated in the $T$-dependent magnetic susceptibility curve (Figure 1c). The cycloidal spin order gives rise to the spontaneous polarization via the inverse Dzyaloshinskii–Moriya interaction or the spin-dependent metal-ligand hybridization.[20-24] As a result, $NiI_2$ has the multiferroic ground state with the FE order coexisting with the cycloidal AFM order. The magnetoelectric coupling was successfully demonstrated by Kurumaji *et al*., reporting efficient control of spontaneous polarization by applying a magnetic field.[19] Due to the triangular lattice symmetry, the cycloidal order can have six possible orientations of the $q$-vector with $q\sim(0.138, 0, 1.457)$ as one of them.[18,19] Accordingly, there should be six FE domains with the in-plane polarization component directed perpendicularly to $q_{in}$, the in-plane component of the $q$-vector, as indicated in Fig. 1b.

We employ the optical second-harmonic generation (SHG) technique to monitor the polarization order parameter of the multiferroic state in $NiI_2$. The SHG process, being the lowest order nonlinear optical phenomenon, efficiently arises in the centrosymmetry-broken state from the electric dipole contribution. Hence, the multiferroic phase having spontaneous polarization is expected to exhibit strong and characteristic SHG responses reflecting its point group symmetry.[25-28] Figures 1d and 1e compare the SHG anisotropy patterns obtained at 30 and 70 K for the bulk $NiI_2$ crystal. The SHG intensity is measured with a rotation of the sample azimuth in the normal incidence geometry. Polarization states of the fundamental wave and second-harmonic (SH) wave are parallel (XX). The result at 70 K shows a six-fold anisotropy pattern reflecting the symmetry of the triangular lattice. At 30 K, on the other hand, a two-fold pattern is observed with the signal amplitude enhanced by orders of magnitude.

We attribute this considerable enhancement of the SHG intensity at 30 K to the inversion symmetry breaking in the FE state. As proportional to the square of electric polarization,[26,29] the SHG intensity displayed in Figure 1f exhibits an abrupt increase near 58 K upon cooling, which corresponds to the transition temperature to the helimagnetic state. Considering that the FE polarization in $NiI_2$ arises from the cycloidal spin order,[19] these results demonstrate that the SHG responses allow us to trace the $T$-dependent development of the FE order parameter and, conversely, the cycloidal spin order. In particular, the polar axis in the FE state or the $q$-vector of the cycloidal spin order can be easily determined with respect to the crystallographic axis from the SHG anisotropy pattern, as discussed below in detail.

We now turn our attention to examine how the multiferroic phase transition varies with a reduction of the thickness of $NiI_2$. We obtained few-layer $NiI_2$ flakes by mechanical exfoliation and transferred them onto $SiO_2$ (285 nm)/Si substrates. Figures 2a and b display optical images of atomically thin $NiI_2$ flakes, whose thicknesses range from a monolayer 1L to $t$=41 nm. We estimated each flake's thickness by the optical contrast (Figure 2d) and atomic force microscopy measurement; Figure 2e shows representative line profiles for 1L, 2L, and 3L flakes where the single NiI$_2$ layer's thickness is about 0.7 nm.[18,30] Figure 2c presents the SHG mapping result taken at 100 K for the part of the area shown in Figure 2b. Polarization states are set as XX with the sample azimuth $\phi$=300°. The results show a clear position-dependent contrast demonstrating that the SHG response exhibits a considerable difference depending on the flake thickness and a negligible SHG from the substrate. The corresponding SHG image obtained at 7 K is shown in Supplementary Information S2.

We look into the $T$-dependent variations of rotational anisotropy patterns for several representative samples shown in Figure 3. For $t$=41 nm, the azimuth-dependence shows six-fold anisotropy at high $T$ > 60 K (right-most panel). However, it becomes like a two-fold pattern at low $T$, e.g., 5 K. Such a change in the anisotropy pattern is accompanied by a significant increase in the SHG intensity. Similar to bulk results shown in Figure 1, these behaviors around 60 K can be ascribed to clear signatures of the electric polarization development. As the flake thickness varies from 18 nm to 2L, the $T$-dependent SHG responses' overall behaviors remain the same, but the transition temperature $T_c$ shows a systematic reduction. Recently, Liu *et al.* reported the Raman scattering results on a few-layer NiI$_2$ and showed a similar thickness-dependent change of the magnetic transition temperature of NiI$_2$.[30] These results together demonstrate that the electric polarization develops simultaneously with the helimagnetic order. Namely, the spin-order-driven ferroelectric state would be realized accordingly even in a few-layer NiI$_2$, and the multiferroic state remains the ground state of the atomically thin NiI$_2$ down to 2L. In particular, it is worth emphasizing that the magnetoelectric coupling phenomenon is observed in the few-layer NiI$_2$; the spontaneous polarization exhibits a significant change upon the magnetic field application (Supplementary Information S4). Notably, the monolayer flake does not show any noticeable sign of the polarization development down to 4 K.

To better understand SHG responses in the polar phase, we examine the anisotropic patterns of each flake's SHG results. The SHG signal is usually attributed to the bulk or interface/surface contribution. In determining the origin of the observed SHG responses in NiI$_2$ flakes, we can take a helpful hint from the thickness-dependent SHG signal. As shown in Figure 2f, the SHG intensity $I_{SHG}$ in the paramagnetic/paraelectric state (75 K) increases quadratically with increasing the flake thickness. This indicates that the observed SHG intensity is mainly contributed to by the bulk part. Nevertheless, there is a finite offset in such a quadratic dependence, as shown in the lower panel of Figure 2f. This suggests that the surface or interface contribution cannot be ignored, particularly when the flake is thin enough, namely, thinner than 4L. On the other hand, in the polar state (7 K), the quadratic thickness-dependence is maintained to the bilayer flake with no offset levels observed. This confirms that the bulk has a significant contribution to the observed SHG response in the polar state.

Based on this understanding of the SHG sources, we can now fit the observed SHG anisotropy patterns by considering each phase's crystallographic symmetry. In the paraelectric state, NiI$_2$ has the point group $\bar{3}m$. Since it is centrosymmetric, the bulk electric dipole (ED) contribution should be forbidden, and instead, the bulk electric quadrupole (EQ) contribution is considered as $P_i^{EQ} = \chi_{ijkl} E_j \partial_k E_l$.[31] Here, $P_i^{EQ}$ is the second harmonic polarization that plays a role of the source of the SHG response, and $E_j$ is the electric field component of the fundamental wave. $\chi_{ijkl}$ is the third-order susceptibility component. In the point group $\bar{3}m$, the SHG intensity contributed to by the bulk EQ has the azimuth dependence of $I_{XX}^{2\omega}(\phi) = \left(\chi_{xxzx}^{EQ} \cos(3\phi)\right)^2$, where $\chi_{xxzx}^{EQ}$ is a representative susceptibility component of the point group $\bar{3}m$ (Supplementary Information S6.1). At the surface or interface of NiI$_2$ flakes, the symmetry can be lowered to $3m$, and the ED contribution should be the primary source of the SHG response having the azimuth dependence of $I_{XX}^{2\omega}(\phi) = (\chi_{xxx}^{S-ED} \cos(3\phi))^2$ (Supplementary Information S6.2). Since both the EQ and ED contributions have the same azimuth dependence, we fit the SHG results with a single trigonometric function. Solid lines in Figures 1 and 3 are the fitting curves. From this, it is clear that the six-fold periodicity of the SHG anisotropy can be naturally explained by the bulk EQ and also by the surface/interface ED contributions of the point group $\bar{3}m$ and $3m$, respectively. Notably, the crystallographic axes can be determined from this analysis; two-fold rotation axes of the point group $\bar{3}m$ correspond to the directions with no (minimum) SHG intensity, and accordingly, the crystalline axes [100] and [1$\bar{1}$0] are labeled in Figure 1 (Supplementary Information S7). Note that the point group symmetry in the paraelectric phase remains the same as $\bar{3}m$ irrespective of the thickness of NiI$_2$.

For the FE state, we analyze the SHG anisotropy pattern by considering the bulk ED contribution. The FE phase of the bulk NiI$_2$ has the point group 2,[18,19] and the azimuth-dependence of the SHG intensity is given as $I_{XX}^{2\omega}(\phi) = \frac{1}{16}[A\sin(3\phi) - B\sin(\phi)]^2$ with $A = \chi_{xxy} + \chi_{xyx} - \chi_{yyy} - \chi_{yxx}$ and $B = \chi_{xxy} + \chi_{xyx} - \chi_{yyy} - 3\chi_{yxx}$. It should be noted that all the results cannot be reproduced by using a single-phase with point group 2. Although the results for 2L, 3L, and 4L seem to have a well-defined two-fold periodicity, the single-phase contribution is not enough to fit such results, in particular, of the crossed (XY) polarization configuration (Supplementary Information S8.1). Instead, we consider contributions from multi-domains, which can have six distinct axes of spontaneous polarization. The SHG mapping image exhibits a highly inhomogeneous distribution of the SHG intensity (Fig. S2f in Supplementary Information). We adjust each polarization domain's relative contribution by varying nonzero susceptibility components and successfully fit all the results obtained in the polar state. For the bulk NiI$_2$, the FE polarization is aligned along the two-fold rotation axis, which in our fitting analyses is set along [100], i.e., $\phi$=30° and the other two axes along $\phi$=90 and 150° (Supplementary Information S8.2). These results are consistent with the polarization directions reported previously.[19] For the thin flake NiI$_2$, on the other hand, the successful fit demands the electric polarization along $\phi$=0, 60, and 120° (Supplementary Information S8.3). This suggests the symmetry point group *m* instead of 2, and it is reminiscent of the other type of the helimagnetic state with the *q*-vector along [100] or its family directions. Note that the helimagnetic state with the *q*-vector along [100] is the magnetic ground state of the bulk NiBr$_2$.[32] Recently, Amoroso *et al*. theoretically demonstrated that the monolayer NiI$_2$ could have a similar magnetic ground state with NiBr$_2$.[33] According to the simple Heisenberg Hamiltonian, two helimagnetic states with such different sets of the *q*-vector have a tiny energy difference, namely <0.1 meV/Ni$^{2+}$.[34] (Supplementary Information S9) Hence, a perturbation to the original Hamiltonian, such as an interlayer spin interaction, may lead to the stabilization of one magnetic state than the other.

We can address the essential points about the magnetic interactions from this analysis result. Whereas three domains are required to reproduce experimental results, we find that all three domains seem to be populated with similar weights having no preferred orientation for the polarization axis (Supplementary Information S8.4). The SHG anisotropy patterns of NiI$_2$ flakes exhibit weak but residual two-fold anisotropy at $T>T_c$ (Fig. 3). Hence the flakes studied in this work seem to be strained to some extent as often observed for flakes of vdW materials which are prepared by a mechanical exfoliation followed by a transfer process onto the substrate[35-37]. Nevertheless, considering the population of three (six) domains with a similar density, we can conclude that such strain effect is not distinguishable, at least in the polar state. This means that there is no apparent easy-axis anisotropy, and hence our system can be considered the simple Heisenberg system.

We now examine the $T$-dependent SHG intensity variations in more detail. Figure 4a shows the $T$-dependent SHG intensity of six $NiI_2$ flakes with different thicknesses. We can easily assign the polar transition temperature $T_c$ from an onset temperature where the SHG intensity increases with a temperature decrease. Figure 4b displays $T_c$ with a variation of the flake thickness. (The $T$-dependent $I_{SHG}$ for 8L and 11L are shown in Supplementary Information S10.) Transition temperature decreases from 58 K for the bulk to about 20 K for 2L. Such a considerable reduction of the transition temperature with a reduction of the flake thickness is also confirmed from the Monte Carlo simulation considering a simple spin Hamiltonian (Supplementary Information S9). This result indicates that the interlayer coupling should play an essential role in determining the magnetic state of $NiI_2$.[38] This behavior may also be understood by the Mermin-Wagner theorem. As aforementioned, our $NiI_2$ flakes endow the Heisenberg magnetic Hamiltonian with no meaningful sign of the easy-axis Ising anisotropy. Accordingly, the magnetic order instability becomes more considerable with a reduction of the flake thickness. It possibly leads to the absence of the long-range magnetic order in the two-dimensional limit. As our finding is only valid in the temperature range down to about 4 K (Figure 3), it will be interesting to examine whether the MF transition would appear at the lower temperature.

Finally, we discuss the robustness of the MF state even with a reduction of the flake thickness. Considering that $I_{SHG}$ is proportional to the square of the electric polarization, $\sqrt{I_{SHG}}$ normalized by the number of layers, displayed in the inset of Figure 4b, confirms that the order parameter strength appears with almost the same amplitude for 2L, 3L, and 4L flakes. This behavior is in strong contrast to the large $T_c$ reduction with a decrease in the flake thickness. The spontaneous polarization in NiI$_2$ arises from the spin order, and its magnitude is in proportion to the $q$-vector.[20-24] Our finding suggests that although the interlayer coupling has noticeable influences on the antiferromagnetic transition temperature and also possibly on the spin texture, the $q$-vector amplitude remains similarly as being less affected by the flake thickness variation.

In summary, we demonstrated the possible persistence of the multiferroicity in atomically thin NiI$_2$ using the optical second-harmonic generation technique. We found that the inversion-symmetry-broken polar state is preserved down to the bilayer NiI$_2$. By analyzing the anisotropy patterns of the SHG responses, we found that polarization domains prevail in both bulk and atomically thin flakes with polarization axes freely chosen along the three (or six) crystallographic axes. As the flake thickness decreases, the transition temperature decreases considerably, implying the interlayer coupling's important role in determining polar and magnetic ground states. The interlayer coupling's effective modulation with a reduction of the flake thickness may lead to the magnetic ground state crossover between different helimagnetic states, as schematically shown in Figure 4c. Our demonstration of possible multiferroicity in the atomically thin vdW material NiI$_2$ will open up a new exciting direction of the magnetoelectric coupling effects in the two-dimensional limit and shed light on enthusiastic efforts to realize multi-functionality in heterostructures of vdW materials.

ASSOCIATED CONTENT

## Supporting Information

Experimental methods, details of the analysis, and supplementary figures. (PDF)

AUTHOR INFORMATION

**Corresponding Author**

Je-Geun Park − Center for Quantum Materials, Seoul National University, Seoul 08826, Republic of Korea; Department of Physics and Astronomy, Seoul National University (SNU), Seoul 08826, Republic of Korea; Institute of Applied Physics, Seoul National University, Seoul 08826, Republic of Korea; E-mail: jgpark10@snu.ac.kr

Jong Seok Lee − Department of Physics and Photon Science, Gwangju Institute of Science and Technology (GIST), Gwangju 61005, Republic of Korea; E-mail: jsl@gist.ac.kr

**Present Addresses**

Chang Jae Roh − *Center for Correlated Electron Systems, Institute for Basic Science (IBS), Seoul 08826, Republic of Korea; Department of Physics and Astronomy, Seoul National University (SNU), Seoul 08826, Republic of Korea*

**Author Contributions**

†H. Ju and Y. Lee contributed equally to this work.

ACKNOWLEDGMENT

The work at GIST was supported by the Ministry of Science, ICT, and Future Planning (Nos. 2015R1A5A1009962, 2018R1A2B2005331). This work at Yonsei University was supported by National Research Foundation (NRF) grants funded by the Korean government (MSIT; grant number 2019R1I1A2A01062306), the SRC program (vdWMRC; grant number 2017R1A5A1014862). The work at CQM and SNU was supported by the Leading Researcher Program of the National Research Foundation of Korea (Grand No. 2020R1A3B2079375). K.-T.K. and K.H.K were supported by National Research Foundation of Korea (NRF) grant funded by the Korea government (MSIT) (NRF- 2019R1A2C2090648). We acknowledge M. S. Song and S. C. Chae for their helps on electric measurements.

REFERENCES

(1) Park, J.-G. Opportunities and challenges of 2D magnetic van der Waals materials: magnetic graphene? *J. Phys. Condens. Matter* **2016,** 28, (30), 301001.

(2) Burch, K. S.; Mandrus, D.; Park, J.-G. Magnetism in two-dimensional van der Waals materials. *Nature* **2018,** 563, (7729), 47-52.

(3) Lee, J.-U.; Lee, S.; Ryoo, J. H.; Kang, S.; Kim, T. Y.; Kim, P.; Park, C.-H.; Park, J.-G.; Cheong, H. Ising-type magnetic ordering in atomically thin $FePS_3$. *Nano Lett.* **2016,** 16, (12), 7433-7438.

(4) Huang, B.; Clark, G.; Navarro-Moratalla, E.; Klein, D. R.; Cheng, R.; Seyler, K. L.; Zhong, D.; Schmidgall, E.; McGuire, M. A.; Cobden, D. H.; Yao, W.; Xiao, D.; Jarillo-Herrero, P.; Xu, X. Layer-dependent ferromagnetism in a van der Waals crystal down to the monolayer limit. *Nature* **2017,** 546, (7657), 270-273.

(5) Fei, Z.; Huang, B.; Malinowski, P.; Wang, W.; Song, T.; Sanchez, J.; Yao, W.; Xiao, D.; Zhu, X.; May, A. F.; Wu, W.; Cobden, D. H.; CHu, J.-H.; Xu, X. Two-dimensional itinerant ferromagnetism in atomically thin $Fe_3GeTe_2$. *Nat. Mater.* **2018,** 17, (9), 778-782.

(6) Kim, K.; Lim, S. Y.; Lee, J.-U.; Lee, S.; Kim, T. Y.; Park, K.; Jeon, G. S.; Park, C.-H.; Park, J.-G.; Cheong, H. Suppression of magnetic ordering in XXZ-type antiferromagnetic monolayer $NiPS_3$. *Nat. Commun.* **2019,** 10, (1), 345.

(7) Kim, K.; Lim, S. Y.; Kim, J.; Lee, J.-U.; Lee, S.; Kim, P.; Park, K.; Son, S.; Park, C.-H.; Park, J.-G.; Cheong, H. Antiferromagnetic ordering in van der Waals 2D magnetic material $MnPS_3$ probed by Raman spectroscopy. *2D Mater.* **2019,** 6, (4), 041001.

(8) Chu, H.; Roh, C. J.; Island, J. O.; Li, C.; Lee, S.; Chen, J.; Park, J.-G.; Young, A. F.; Lee, J. S.; Hsieh, D. Linear Magnetoelectric Phase in Ultrathin $MnPS_3$ Probed by Optical Second Harmonic Generation. *Phys. Rev. Lett.* **2020,** 124, (2), 027601.

(9) Yuan, S.; Luo, X.; Chan, H. L.; Xiao, C.; Dai, Y.; Xie, M.; Hao, J. Room-temperature ferroelectricity in $MoTe_2$ down to the atomic monolayer limit. *Nat. Commun.* **2019,** 10, (1), 1775.

(10) Xue, F.; Hu, W.; Lee, K. C.; Lu, L. S.; Zhang, J.; Tang, H. L.; Han, A.; Hsu, W. T.; Tu, S.; Chang, W. H.; Lien, C.-H.; He, J.-H.; Zhang, Z.; Li, L.-J.; Zhang, X. Room-Temperature Ferroelectricity in Hexagonally Layered α-$In_2Se_3$ Nanoflakes down to the Monolayer Limit. *Adv. Funct. Mater.* **2018,** 28, (50), 1803738.

(11) Higashitarumizu, N.; Kawamoto, H.; Lee, C.-J.; Lin, B.-H.; Chu, F.-H.; Yonemori, I.;

Nishimura, T.; Wakabayashi, K.; Chang, W.-H.; Nagashio, K. Purely in-plane ferroelectricity in monolayer SnS at room temperature. *Nat. Commun.* **2020,** 11, (1), 2428.

(12) Zhong, D.; Seyler, K. L.; Linpeng, X.; Wilson, N. P.; Taniguchi, T.; Watanabe, K.; McGuire, M. A.; Fu, K.-M. C.; Xiao, D.; Yao, W.; Xu, X. Layer-resolved magnetic proximity effect in van der Waals heterostructures. *Nat. Nanotechnol.* **2020,** 15, (3), 187-191.

(13) Mukherjee, A.; Shayan, K.; Li, L.; Shan, J.; Mak, K. F.; Vamivakas, A. N. Observation of site-controlled localized charged excitons in $CrI_3$/$WSe_2$ heterostructures. *Nat. Commun.* **2020,** 11, (1), 5502.

(14) Wu, Y.; Zhang, S.; Zhang, J.; Wang, W.; Zhu, Y. L.; Hu, J.; Yin, G.; Wong, K.; Fang, C.; Wan, C.; Han, X.; Shao, Q.; Taniguchi, T.; Watanabe, K.; Zang, J.; Mao, Z.; Zhang, X.; Wang, K. L. Néel-type skyrmion in $WTe_2$/$Fe_3GeTe_2$ van der Waals heterostructure. *Nat. Commun.* **2020,** 11, (1), 3860.

(15) Huang, X.; Li, G.; Chen, C.; Nie, X.; Jiang, X.; Liu, J.-M. Interfacial coupling induced critical thickness for the ferroelectric bistability of two-dimensional ferromagnet/ferroelectric van der Waals heterostructures. *Phys. Rev. B* **2019,** 100, (23), 235445.

(16) Yin, L.; Parker, D. S. Out-of-plane magnetic anisotropy engineered via band distortion in two-dimensional materials. *Phys. Rev. B* **2020,** 102, (5), 054441.

(17) Zhao, Y.; Zhang, J. J.; Yuan, S.; Chen, Z. Nonvolatile electrical control and heterointerface-induced half-metallicity of 2D ferromagnets. *Adv. Funct. Mater.* **2019,** 29, (24), 1901420.

(18) Kuindersma, S. R.; Sanchez, J. P.; Haas, C. Magnetic and structural investigations on $NiI_2$ and $CoI_2$. *Physica B+C* **1981,** 111, (2-3), 231-248.

(19) Kurumaji, T.; Seki, S.; Ishiwata, S.; Murakawa, H.; Kaneko, Y.; Tokura, Y. Magnetoelectric responses induced by domain rearrangement and spin structural change in triangular-lattice helimagnets $NiI_2$ and $CoI_2$. *Phys. Rev. B* **2013,** 87, (1), 014429.

(20) Katsura, H.; Nagaosa, N.; Balatsky, A. V. Spin current and magnetoelectric effect in noncollinear magnets. *Phys. Rev. Lett.* **2005,** 95, (5), 057205.

(21) Sergienko, I. A.; Dagotto, E. Role of the Dzyaloshinskii-Moriya interaction in multiferroic perovskites. *Phys. Rev. B* **2006,** 73, (9), 094434.

(22) Jia, C.; Onoda, S.; Nagaosa, N.; Han, J. H. Bond electronic polarization induced by spin. *Phys. Rev. B* **2006,** 74, (22), 224444.

(23) Jia, C.; Onoda, S.; Nagaosa, N.; Han, J. H. Microscopic theory of spin-polarization coupling

in multiferroic transition metal oxides. *Phys. Rev. B* **2007,** 76, (14), 144424.

(24) Arima, T.-h. Ferroelectricity induced by proper-screw type magnetic order. *J. Phys. Soc. Japan* **2007,** 76, (7), 073702-073702.

(25) Lilienblum, M.; Lottermoser, T.; Manz, S.; Selbach, S. M.; Cano, A.; Fiebig, M. Ferroelectricity in the multiferroic hexagonal manganites. *Nat. Phys.* **2015,** 11, (12), 1070-1073.

(26) Lottermoser, T.; Meier, D.; Pisarev, R. V.; Fiebig, M. Giant coupling of second-harmonic generation to a multiferroic polarization. *Phys. Rev. B* **2009,** 80, (10), 100101.

(27) Roh, C. J.; Lee, J. H.; Kim, K.-E.; Yang, C.-H.; Lee, J. S. Deterministic domain reorientations in the $BiFeO_3$ thin film upon the thermal phase transitions. *Appl. Phys. Lett.* **2018,** 113, (5), 052904.

(28) Sherstyuk, N. E.; Mishina, E. D.; Lavrov, S. D.; Buryakov, A. M.; Marchenkova, M. A.; Elshin, A. S.; Sigov, A. S. Optical second harmonic generation microscopy for ferroic materials. *Ferroelectrics* **2015,** 477, (1), 29-46.

(29) Meier, D.; Leo, N.; Yuan, G.; Lottermoser, T.; Fiebig, M.; Becker, P.; Bohatý, L. Second harmonic generation on incommensurate structures: The case of multiferroic $MnWO_4$. *Phys. Rev. B* **2010,** 82, (15), 155112.

(30) Liu, H.; Wang, X.; Wu, J.; Chen, Y.; Wan, J.; Wen, R.; Yang, J.; Liu, Y.; Song, Z.; Xie, L. Vapor Deposition of Magnetic Van der Waals $NiI_2$ Crystals. *ACS Nano* **2020,** 14, (8), 10544-10551.

(31) Roh, C. J.; Jung, M. C.; Kim, J. R.; Go, K. J.; Kim, J.; Oh, H. J.; Jo, Y. R.; Shin, Y. J.; Choi, J. G.; Kim, B. J.; Noh, D. Y.; Choi, S.-Y.; Noh, T. W.; Han, M. J.; Lee, J. S. Polar Metal Phase Induced by Oxygen Octahedral Network Relaxation in Oxide Thin Films. *Small* **2020,** 16, (40), 2003055.

(32) Tokunaga, Y.; Okuyama, D.; Kurumaji, T.; Arima, T.; Nakao, H.; Murakami, Y.; Taguchi, Y.; Tokura, Y. Multiferroicity in $NiBr_2$ with long-wavelength cycloidal spin structure on a triangular lattice. *Phys. Rev. B* **2011,** 84, (6), 060406.

(33) Amoroso, D.; Barone, P.; Picozzi, S. Spontaneous skyrmionic lattice from anisotropic symmetric exchange in a Ni-halide monolayer. *Nat. Commun.* **2020,** 11, (1), 5784.

(34) Rastelli, E.; Tassi, A.; Reatto, L. Non-simple magnetic order for simple Hamiltonians. *Physica B+C* **1979,** 97, (1), 1-24.

(35) Deng, S.; Sumant, A. V.; Berry, V. Strain engineering in two-dimensional nanomaterials beyond graphene. *Nano Today* **2018,** 22, 14-35.

(36) Mennel, L.; Paur, M.; Mueller, T. Second harmonic generation in strained transition metal

dichalcogenide monolayers: $MoS_2$, $MoSe_2$, $WS_2$, and $WSe_2$. *APL Photonics* **2019,** 4, (3), 034404.

(37) Mennel, L.; Furchi, M. M.; Wachter, S.; Paur, M.; Polyushkin, D. K.; Mueller, T. Optical imaging of strain in two-dimensional crystals. *Nat. Commun.* **2018,** 9, (1), 1-6.

(38) Huang, F.; Kief, M.; Mankey, G.; Willis, R. Magnetism in the few-monolayers limit: A surface magneto-optic Kerr-effect study of the magnetic behavior of ultrathin films of Co, Ni, and Co-Ni alloys on Cu (100) and Cu (111). *Phys. Rev. B* **1994,** 49, (6), 3962.

Figures

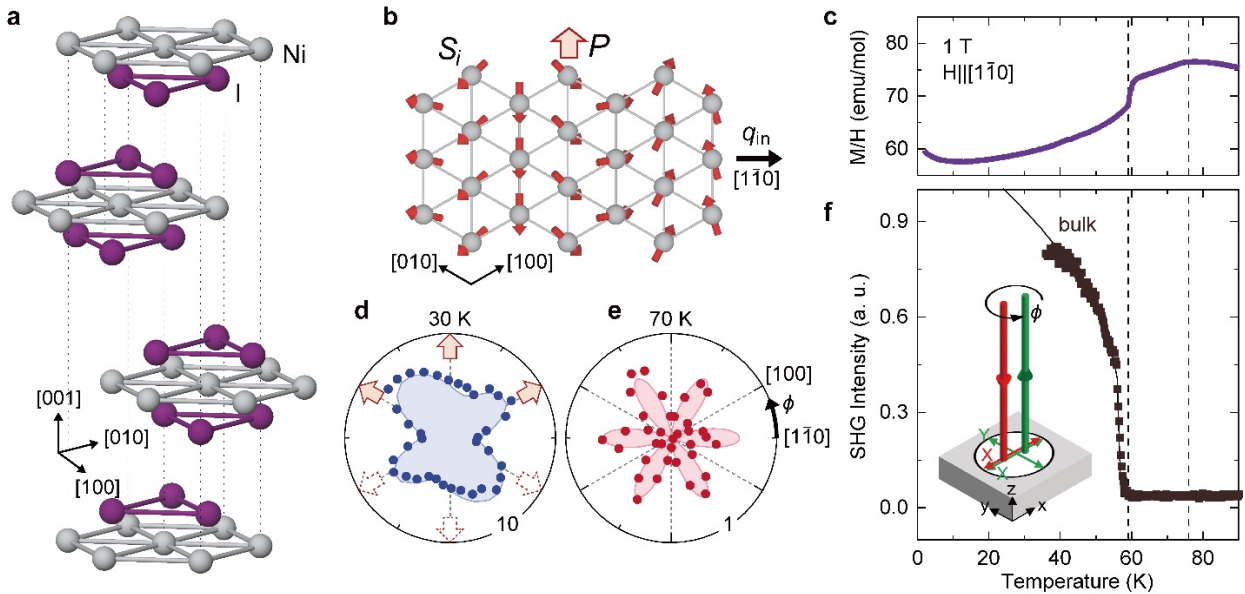

**Figure 1.** Bulk NiI$_2$ crystals with the multiferroic ground state. (a) The layered triangular crystalline structure of NiI$_2$. Grey and purple balls represent nickel and iodine atoms, respectively. (b) Cycloidal order of magnetic moments in a single layer NiI$_2$. Red arrows denote spin $S_i$ of Ni atoms having the cycloidal order with the modulation vector $q_{in}$ along [1$\bar{1}$0]. The spin-order-induced polarization $P$ is also indicated. (c) Temperature-dependent magnetic susceptibility of the bulk NiI$_2$ with a magnetic field applied along [1$\bar{1}$0]. Dashed vertical lines indicate the magnetic transition to the collinear antiferromagnet at 76 K and the helimagnet at 58 K. (d), (e) Rotation anisotropy patterns of the second-harmonic generation (SHG) intensity for the bulk NiI$_2$ obtained at 30 K and 70 K. Symbols are experimental results and solid lines are fitting curves. Block arrows in (d) denote the polarization directions in the multi-domain ferroelectric state. Directions of two crystallographic axes are also indicated in (e). (f), Temperature-dependent SHG intensity at the selected azimuth angle. A solid line is a guide to the eye. The inset shows a schematic for the normal-incidence SHG measurement where polarization states of the fundamental and second-harmonic waves are set to be parallel (XX) or crossed (XY).

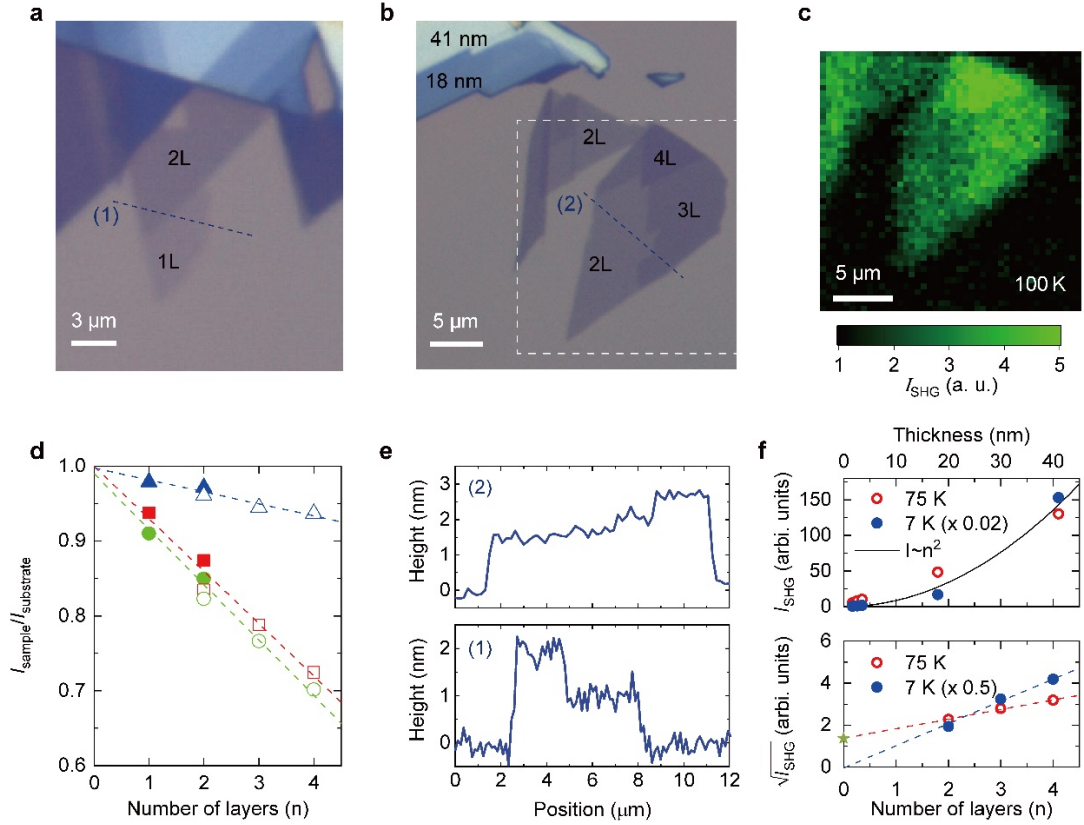

**Figure 2.** Thickness characterization of NiI$_2$ flakes. (a), (b) Optical images of a few-layer NiI$_2$. (c) Second-harmonic generation signal mapping image at 100 K for the squared area marked in (b). (d) Thickness-dependent optical contrasts for the red, green, and blue colors for the optical images shown in (a) and (b). Lines are linear fits. (e) Line profiles of the atomic force microscope results along the lines (1) and (2) indicated in (a) and (b). Here, the flakes with 1L, 2L, and 3L are distinguished. (f), Thickness-dependent SHG intensity. The red open symbol indicates data taken at 75 K, while the blue closed symbol does data taken at 7 K. While both results scale well with the quadratic dependence on the thickness, the results at 75 K have the finite offset, which is indicated with the green star at the vertical-axis intercept in the lower panel.

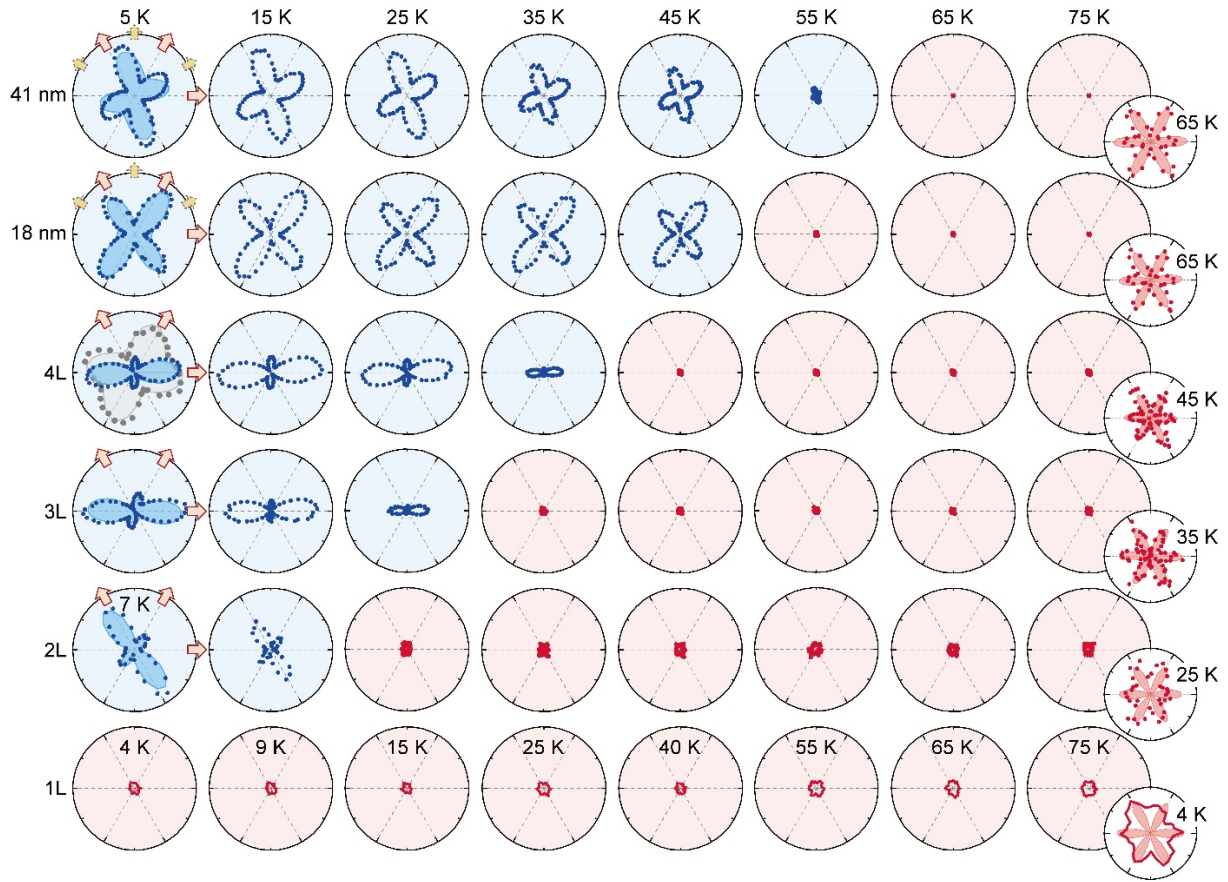

**Figure 3.** Rotational anisotropy of the second-harmonic generation (SHG) pattern for several NiI$_2$ flakes at selected temperatures. The layer number of NiI$_2$ flakes ranges from 1L to about 60L, corresponding to the thickness $t$=41 nm. Measurement temperatures are given differently for 1L and 2L (only at the lowest temperature, 7 K). Two different background colors are used to highlight the transition to the polar state. The final fitting results are shown at the left-most and right-most figures with filled patterns. For 4L, additional data obtained is displayed with a gray color. Block arrows in the left-most figure denote the polarization directions in the multi-domain state, producing experimental results. Although polarization axes are decisively determined for 2L, 3L, and 4L, two sets of polarization axes can be considered to fit the results for $t$=18 and 41 nm.

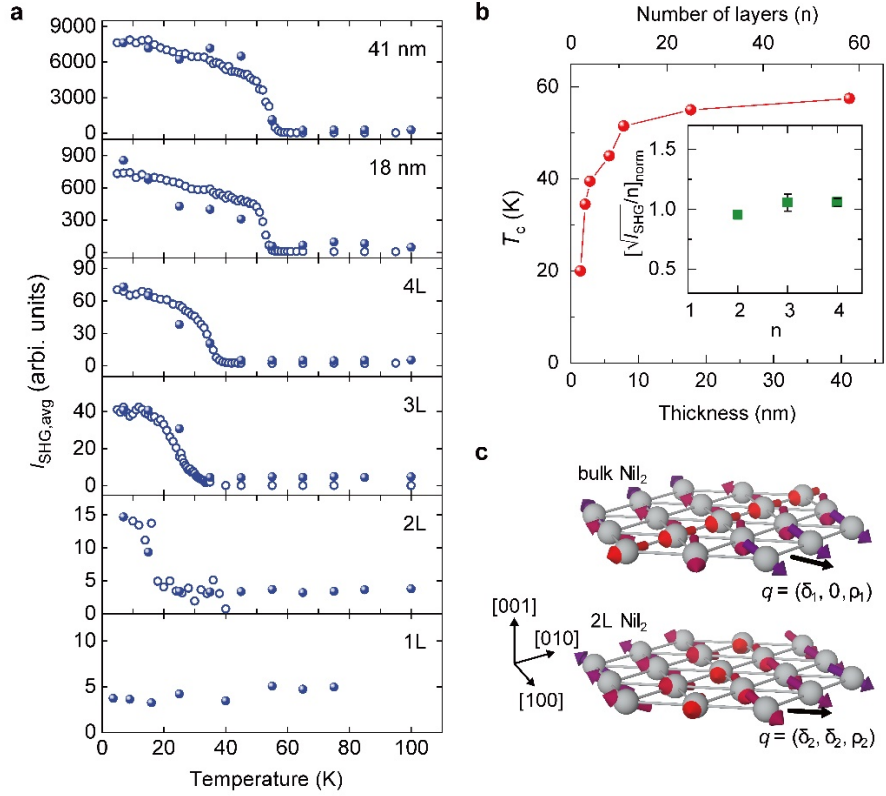

**Figure 4.** Thickness-dependent evolution of the multiferroic state of NiI$_2$ flakes. (a) Temperature-dependent second-harmonic generation (SHG) intensity. Solid and open symbols are obtained using different detection schemes, namely, photon counting and modulation-based (lock-in) detection. Also note that the two data sets were obtained at a time interval of two months, and samples were kept in an Ar gas during the storage. (b) Thickness-dependent transition temperature to the polar state, taken from the onset of the SHG intensity shown in (a). The inset shows that $\sqrt{I_{\text{SHG}}}/n$, being in proportion to the order parameter strength for the polar state, appears with almost the same amplitude irrespective of the thickness. The finite thickness effect is also considered for the fair comparison (Supplementary Information S11). (c) The two-dimensional spin structure proposed for the 2L NiI$_2$. Ball and arrow denote Ni ion and its spin, respectively. For the 2L NiI$_2$, the propagation $q$-vector in the helimagnetic state is defined along the Ni-Ni bonding axis.